\documentclass[final,5p,fleqn]{elsarticle}
\usepackage{amsmath,amssymb}
\usepackage{epsfig}
\usepackage{color}
\usepackage{graphicx}

\journal{Physics Letters B}

\begin{document}

\begin{frontmatter}



\title{Sensitivity of transfer cross sections to the bound-state wave functions}


\author{Shubhchintak}
\ead{shubhchintak@ulb.ac.be} 
\author{P. Descouvemont\fnref{label2}}
\fntext[label2]{Directeur de Recherches FNRS}
\ead{pdesc@ulb.ac.be} 

\address{Physique Nucl\'{e}aire Th\'{e}orique et Physique Math\'{e}matique, C. P. 229, Universit\'{e} Libre de Bruxelles (ULB), B 1050 Brussels, Belgium}

\begin{abstract}
We test the sensitivity of transfer reactions to the bound state wave functions within a distorted wave Born approximation formalism. Using supersymmetric transformations, we remove the Pauli-forbidden states from the two-body potentials and generate an equivalent supersymmetric partner. 
Wave functions from these potentials have the same asymptotics, but they differ in the nuclear interior. This allows us to study the influence of the nuclear interior on transfer cross sections.
We apply the calculations to the $^{16}$O($d, p$)$^{17}$O and $^{12}$C($^7$Li, $t$)$^{16}$O reactions, which are typical examples of nucleon and $\alpha$ transfer, respectively. The spectroscopic factors for $^{17}$O are decreased by about
30\% when using supersymmetric potentials. For $^{16}$O, the differences are smaller. However, we show that ambiguities exist
in the determination of the spectroscopic factors, due to the choice of the angular range where the fit is performed.
\end{abstract}


\end{frontmatter}


Nuclear transfer reactions have many applications, from  the nuclear structure to the field of astrophysics. Due to their sensitivity to the overlap functions in the initial and final channels, from more than six decades they are being used as an important tool to investigate the nuclear structure \cite{Sa90,Gl04,OII69,Th88}. With the advent of radioactive ion beam facilities, their applications were further extended to drip-line nuclei by invoking the inverse kinematics (see, for example, Ref.\ \cite{THO05}). In particular, from the angular distribution of the cross sections, one can identify the angular momentum ($\ell$) of the residual nucleus and further combining it with additional information (reaction with polarized beams) spin-parity ($J^\pi$) of a particular residual state can be figured out. Furthermore, comparing the measured angular distribution with the calculated ones, spectroscopic factors or asymptotic normalization coefficients (ANC) (defined as the amplitude of the tail of the overlap function) can be obtained for the final nucleus. A recent review of $(d,p)$ reactions is provided in Ref.\ \cite{TJ20}.

In the field of nuclear astrophysics, where one mostly deals with low energy reactions which are often difficult (sometime impossible) to measure directly, transfer reactions are used as one of the indirect tools to extract spectroscopic information such as energies, widths or spins, etc., needed to study such reactions \cite{BAR16,BER16,TRI14}. For example, cross sections of low-energy radiative capture reactions are mainly determined by the ANCs, provided the process is peripheral. Some recent studies are focused on extracting neutron capture cross sections for the $r$-process nuclei using ($d, p$) reactions \cite{THO05,AKR08,BER19}. Theoretical estimates based on statistical models such as the Hauser-Feshbach theory (which are often used for neutron capture cross sections calculations) are usually not reliable due to the low level densities for most of the nuclei involved in r-process paths. Therefore, ($d, p$) reactions are being considered as an alternative tool to extract the desired neutron capture cross sections. Similarly, $\alpha$ transfer reactions such as ($^7$Li, $t$) and ($^6$Li, $d$) have been used to study the important astrophysical reactions like  $^{12}$C($\alpha, \gamma$)$^{16}$O, $^{13}$C($\alpha, n$)$^{16}$O, $^{22}$Ne($\alpha, \gamma$)$^{26}$Mg, (for details see for example the review article \cite{BAR16}).

At the simplest, one can consider transfer as a three-body reaction, and with that picture several theoretical models using different frameworks and approximations have been developed in the literature. The standard and the most widely used framework is `the distorted wave Born approximation (DWBA)' \cite{Sa83}, where one sees transfer as a one-step process, weak enough to be treated as a first order perturbation. The three-body wave function in the entrance channel is approximated by a product of the internal and relative wave functions. Other methods such as the adiabatic method \cite{JS70}, the continuum discretized coupled-channel (CDCC) method \cite{AIK87} and the Faddeev method \cite{De13} are more advanced than standard DWBA and take into account the genuine three-body wave function along with higher-order effects such as breakup. 
But due to the complexities of the CDCC and Faddeev's method, since a long time DWBA have been used for the analysis of the experimental data. 
Of course, with the recent advancements in reaction theories, these methods are now also being adopted in several studies involving deuteron \cite{De13,MNJ09, NNJ10}, especially the adiabatic method which treats the breakup effects in an approximated way \cite{SJA13,WPC19}. However, because of its simplicity, DWBA is still being used in many data analysis, mainly to extract the spectroscopic factors or ANCs [see for example \cite{JRG20,HWK20}].


Spectroscopic factors extracted using the DWBA often contain uncertainties due to the ambiguities in the optical and bound state potentials. To minimize such uncertainties, phenomenological optical potentials which fit the elastic scattering data at the reaction energies are preferred \cite{KLN18}. Simultaneous measurements of the elastic and the transfer cross section also serve as ideal choice especially for the unstable nuclei where elastic scattering data do not always exist. In such cases analysis of the elastic scattering feeds potential to the transfer calculations \cite{SSB77,SJB12}. Similarly, for the bound states, two body potentials such as Woods-Saxon (WS) or Gaussian type are normally adopted with their parameters adjusted to reproduce the experimental binding energies. But again, variations in these parameters influence the cross sections and hence the extracted spectroscopic factors (see for example 
Refs.\ \cite{PNM07,YC18}). Furthermore, two-body potentials constructed for these many-body systems contain Pauli-forbidden states, which are unphysical but simulate the missing antisymmetrization effects. 

Supersymmetric quantum mechanics, on the other hand, provides a way to remove these deep states \cite{SUK85,BAY87} by transforming a potential to its supersymmetric (SUSY) partner, which, at positive energies, gives the same phase shifts. The resulted $\ell$-dependent potentials are characterized by a repulsive core and are usually shallow \cite{BAY87} at intermediate distances. By virtue of this, wave functions in the nuclear interior also differ and possess a nodeless structure. In spite of being different in the nuclear interior, wave functions generated by these shallow (SUSY) and deep potentials give similar spectroscopic properties, such as root mean square (r.m.s) radii \cite{RVB96}. 
 
In this paper, using a Woods-Saxon and its SUSY partner, we study the region of the wave function being probed by the transfer reactions in the DWBA formalism. Subsequently, we compare the respective spectroscopic factors extracted by fitting the calculated cross sections to the available experimental data for the reactions of our interest.
In particular, we consider the $^{16}$O($d, p$)$^{17}$O and $^{12}$C($^7$Li, $t$)$^{16}$O reactions, specific cases of $n$ and $\alpha$ transfer on stable targets. Throughout the paper we will use the name `deep potential' for WS and `SUSY potential' for its supersymmetric partner. Note that, using SUSY potentials the antisymmetrization is taken into account only within the projectile and/or in the residual nucleus but not in the projectile + target system.

We use the DWBA formalism of transfer reactions developed in Ref. \cite{SP19} where we have utilized the $R$-matrix \cite{PDB10} and Lagrange-mesh \cite{BAYE15} methods which, apart from simplifying the calculations, also lead to fast and accurate numerical computations. 
In the DWBA formalism the scattering matrix for the stripping reaction $A(a+c) + T \rightarrow B(T+c)+ a$, where a cluster $c$ is transferred from the projectile $A$ to the target $T$ (spin-less in our case) leading to the formation of nucleus $B$ and cluster $a$ in the final channel, can be simply written as

\begin{eqnarray}
U^{J\pi}_{\alpha \beta} =  
-\frac{i\sqrt{S_A S_B}}{\hbar}\big\langle\chi^{J\pi}_{L_B}(\pmb{R}')\Phi^{I_B}_{\ell_B}(\pmb{r}_B)|\mathcal{V}|\chi^{J\pi}_{L_A}(\pmb{R})\Phi^{I_A}_{\ell_A}(\pmb{r}_A)\big\rangle,
\label{a0}
\end{eqnarray} 
where $\Phi^{I_k}_{\ell_k}$ is the two-body bound state wave function of the nucleus $k$, having orbital angular momentum $\ell_k$ which couple with the spins of the core and the valence particle to give the total angular momentum $I_k$ \cite{SP19}.
$\chi_{L_A}(\pmb{R})$ and $\chi_{L_B}(\pmb{R}')$ are the scattering wave functions for the relative motion of nuclei $A-T$ and $B-a$ in the initial and final channels with angular momenta $L_A$ and $L_B$, respectively. $J$ is the total angular momentum and $\pi$ is the parity.  $S_A$ and $S_B$ are the spectroscopic factors. In the above definition, labels $\alpha$ and $\beta$ stand for ($L_A,\ell_A,I_A$) and ($L_B,\ell_B,I_B$), respectively, and $\pmb{r}_A$, $\pmb{r}_{B}$, $\pmb{R}$ and $\pmb{R'}$ are the relative coordinates between the center of mass of the respective clusters. 

In Eq.\ (\ref{a0}), the interaction $\mathcal{V} = V_{ac}+U_{aT}-U_{aB}$ in the post form or $\mathcal{V} = V_{Tc}+U_{aT}-U_{AT}$ in the prior form, where, $V_{ik}$ is the binding potential between $i-k$, $U_{aT}$ is the core-core optical potential and $U_{AT}$, $U_{aB}$ are the scattering potentials in the initial and final channels, respectively. The later two terms in both  definitions are called {\it remnant terms} and they often appear similar. They are sometimes neglected to simplify the calculations.  
Though this approximation is valid for heavy targets, however, the remnant term can contribute in reactions involving 
light nuclei (see, for example, the discussion in
Ref.\ \cite{SP19}). In the DWBA, the post and the prior forms give same results provided the remnant terms are included  \cite{TN09}.

Eq. (\ref{a0}) can further be written as
\begin{eqnarray}
U^{J\pi}_{\alpha \beta} =  -\frac{i\sqrt{S_A S_B}}{\hbar}\int \chi^{J\pi}_{L_A}(R)K^J_{\alpha \beta}(R, R')\chi^{J\pi}_{L_B}(R')R R' dR dR',
\label{a1}
\end{eqnarray}
where the transfer kernel $K(R, R')$ is defined by
\begin{eqnarray}
K^{J \pi}_{\alpha \beta}(R,R')&=&\bigg\langle\Big[Y_{L_A}(\Omega)\otimes \Phi^{I_A}_{\ell_A}({\bf r}_A)\Big]^J \vert \mathcal{V} \nonumber\\
& &\vert \Big[Y_{L_B}(\Omega')\otimes \Phi^{I_B}_{\ell_B}({\bf r}_B)\Big]^J\bigg\rangle. 
\label{a2}
\end{eqnarray}
From the scattering matrices (\ref{a1}), the transfer cross sections are easily determined \cite{SP19,PDB10}.  
We have verified the equivalence of the post and prior forms in all our calculations, but in the following we mainly discuss our results calculated under the aegis of post form DWBA.

All  potentials used here are the same as in Ref.\ \cite{SP19} and they were adopted from Refs. \cite{CHR74,OHR12,PDM15,KD03,SUD73,GH73}. For the simplicity we have dropped the spin-orbit term in the scattering potentials.
We then apply a SUSY transformation to the bound-state potentials to remove the states associated with Pauli-forbidden state. In fact, in the supersymmetric quantum mechanics one can transform a Hamiltonian 
\begin{eqnarray}
H_0 = -\frac{\hbar^2}{2\mu}\frac{d^2}{dr^2} +V^{\ell j}_0(r) 
\end{eqnarray}
to its partner Hamiltonian $H_2$ having the identical bound-state energy spectrum except for the lowest states which are suppressed \cite{SUK85}. This can be achieved by performing couple of transformations \cite{BAY87}, which change the potential to $V^{\ell j}_2$, given by
\begin{eqnarray}
V^{\ell j}_2 = V^{\ell j}_0-\frac{\hbar^2}{\mu}\frac{d^2}{dr^2} \,\log \int_0^r |u_0(r)|^2 dr,
\label{a3}
\end{eqnarray}
where $\mu$ is the reduced mass, $\ell$ and $j$ stand for the orbital angular momentum and spin, respectively. Note that potential $V^{\ell j}_0$ contains Coulomb, centrifugal and spin-orbit terms. In our calculations, the nuclear part is chosen as
a Woods-Saxon potential ($V_{ac},V_{Tc}$). In Eq.\ (\ref{a3}), $u_0(r)$ is the radial part of the bound-state wave function obtained with the potential $V_0^{\ell j}$. With repeated applications of this procedure one can therefore eliminate all the unphysical states from the spectrum of the parent Hamiltonian. The modified potential $V^{\ell j}_2$ depends on angular momentum, even
if the initial nuclear potential in $V_0$ does not.

The number of forbidden states is given by a parity dependent limit $N$, so that the states having principle quantum number $2n + \ell$ less than $N$ are forbidden. In Table \ref{tab1}, we give the number of forbidden states for various states in the systems considered here. 
Note that there is no forbidden state in the potential used for the ground state of the deuteron and of $^{17}$O. 
For the second part of our calculation, we use Eq.\ (\ref{a3}) to construct SUSY partner potentials $V_2^{\ell j}$. The potentials $V_2^{\ell j}$ possess a $1/r^{2}$ singularity in the origin and are shallow in nature \cite{BAY87}.

\begin{table}[ht]
	\begin{center}
		\caption{Number of Pauli-forbidden states (n) in the different systems considered here.}
		\label{tab1}
		\begin{tabular}{cccc}
			\hline
			\hline
			System     &  state & $\ell$ & n \\
			\hline
			$n+^{16}$O & $1/2^+$ & 0 &   1  \\
			\hline
			$\alpha + t$ & $3/2^-$ & 1 &   1  \\
			\hline
			$\alpha + ^{12}$C & $0_2^+$ & 0 &   4  \\
			\hline
			$\alpha + ^{12}$C & $2_1^+$ & 2 &   3  \\
			\hline
			\hline
		\end{tabular}
	\end{center}
\end{table}

\begin{figure}[ht]
	\includegraphics[ clip,width=0.4\textwidth]{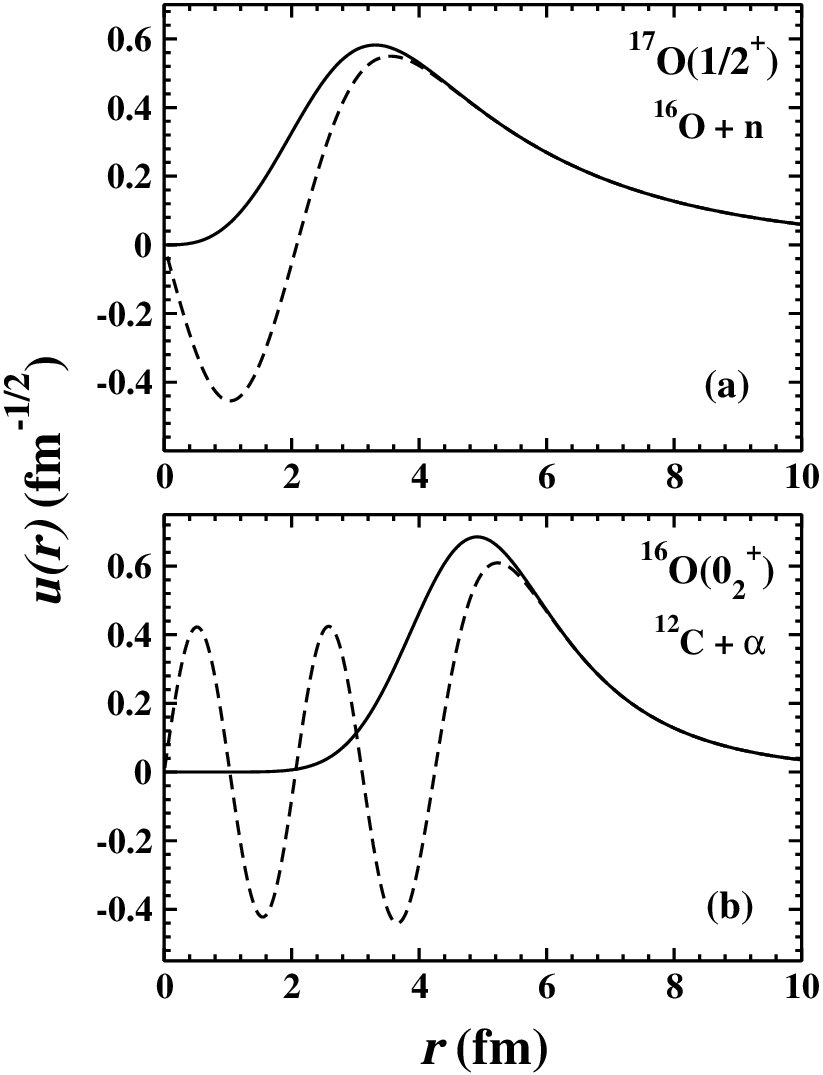}
	\caption{Radial bound-state wave functions of $^{17}$O($1/2^+$) and $^{16}$O($0_2^+$) obtained by solving the Schr\"odinger wave equation with the deep WS potentials (dashed lines) and with the SUSY potentials (solid lines). For details see the text. } 
	\label{wfns}
\end{figure}

In Fig.\ \ref{wfns}, we plot the radial wave functions for the $^{17}$O($1/2^+$, $E_x = 0.86$ MeV) and $^{16}$O($0_2^+$, $E_x = 6.05$ MeV) bound states. Dashed and solid lines represent the wave functions obtained with the WS and with its SUSY partner, respectively. Even though these wave functions differ in the nuclear interior, the r.m.s radii are similar. 

These two different combinations of bound state potentials and wave functions are then used to calculate the transfer cross sections. Note that when using SUSY potential in the matrix element, one has to subtract the centrifugal term from it. This is because the SUSY potential obtained from Eq. (\ref{a3}) are $\ell$ dependent whereas the original matrix potentials are not.
We first illustrate the $^{16}$O($d, p$)$^{17}$O reaction which is the simplest case in present context, with no forbidden bound state in the incident channel potential. As there is no forbidden state in the potential used for the ground state ($5/2^+$), we only consider the striping of a neutron leading to the first excited state ($1/2^+$) of $^{17}$O ($\ell = 0$).

\begin{figure}[ht]
	\includegraphics[ clip,width=0.4\textwidth]{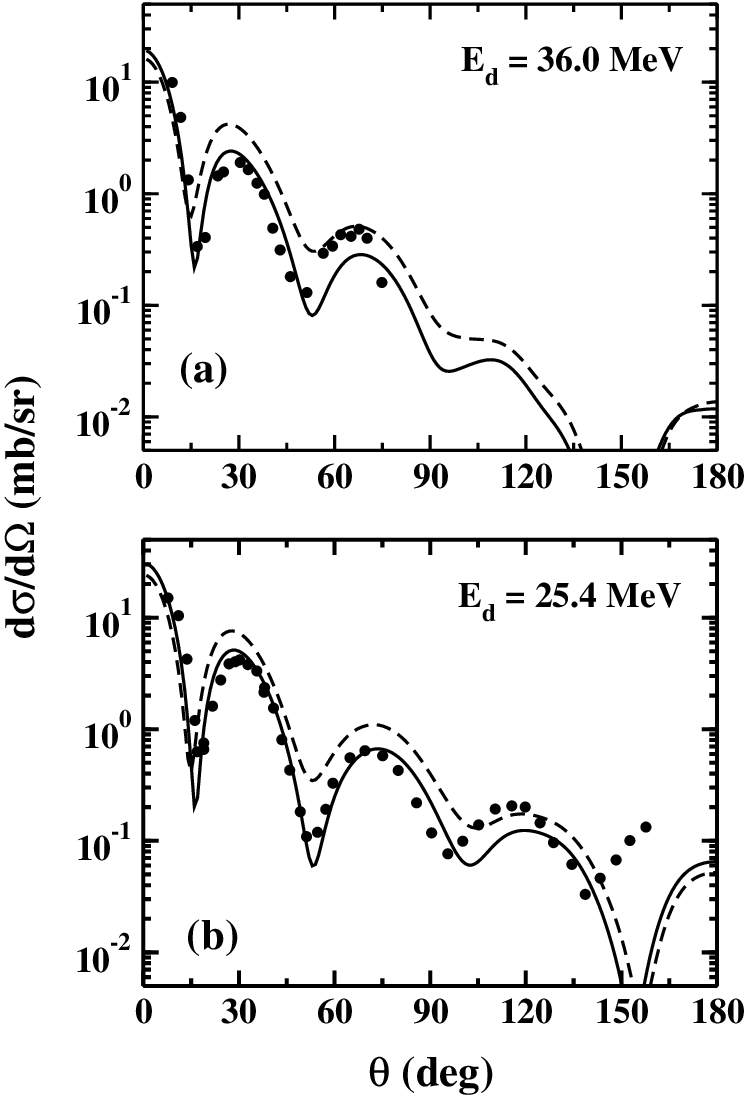}
	\caption{Angular distributions of the $^{16}$O($d, p$)$^{17}$O($1/2^+$) reaction at two different deuteron energies. The dashed and solid lines correspond to the calculations (with SF = 1) using a WS and its SUSY partner for the $^{17}$O($1/2^+$), respectively. Experimental data (solid dots) are taken from Ref.\ \cite{CHR74}. } 
	\label{dp_css}
\end{figure}

In Fig.\ \ref{dp_css}, we plot the the angular distribution for the $^{16}$O($d, p$)$^{17}$O (1/2$^+$) reaction at two different deuteron energies ($E_d = 25.4$ and 36 MeV) and also compare it with the experimental data of Ref.\ \cite{CHR74}. Dashed and solid lines in both panels correspond to the calculations using a WS potential and its SUSY partner for the final bound state. Both  calculations reproduce the shape of the data over a wide angular range (up to $\approx 60^\circ$). However, there is a significant difference in their magnitudes at both energies. Even in the very forward angular range, where transfer reactions are considered as peripheral, both calculations do not give identical results, and the position of the first minima is somewhat different. 

A direct influence of these differences is expected on the spectroscopic factors (SFs), which can be extracted by normalizing the calculations to the experimental data at forward angles. We therefore, use a chi-square minimization procedure to extract the SFs for the (1/2$^+$) excited state of $^{17}$O in both these calculations. 
We fit the data only up to first minima, which is the general procedure to extract the SFs \cite{TN09,LFL04}.
In Table \ref{tab2}, we compare the SFs using both calculations. A decrease of around 20-30\% in the SFs is observed when using a SUSY potentials. The SUSY SFs are more consistent with values derived from a microscopic cluster
model \cite{BT92}. Furthermore, as mentioned in Ref. \cite{CHR74}, around 20\% uncertainties can be expected in the individual SFs due to ambiguities of the optical potentials. Note that our goal here is not to provide the optimal SFs, but it is to study the sensitivity of the SFs to the bound state wave functions. 

\begin{table}[ht]
	\begin{center}
		\caption{Single particle SFs for the $1/2^+$ excited state of $^{17}$O and $\alpha$-SFs for the $0_2^+$ and $2_1^+$ states of $^{16}$O extracted using two different types of bound-state potentials. The SFs for the ground state of $d$ and of
			$^7$Li are taken as 1. Uncertainties in the $^{16}$O SFs arise from different choices of the angular range for the fits (see text).}
		\label{tab2}
		\begin{tabular}{ccccc}
			\hline
			\hline
			Nucleus   & State & Beam energy & SFs  & SFs  \\
			&&(MeV)&(WS)& (SUSY) \\
			\hline
			$^{17}$O & $1/2^+$ & 25.4 & 1.73 &   1.22  \\
						& $1/2^+$ & 36.0 & 2.08 &   1.63  \\
			\hline
			$^{16}$O & $0_2^+$ & 28   & $0.18 \pm 0.04$ & $0.19\pm 0.02$  \\
			$^{16}$O & $0_2^+$ & 34   & $0.24 \pm 0.04$ & $0.25 \pm 0.03$  \\
			\hline
			$^{16}$O & $2_1^+$ & 28   & $0.17 \pm 0.02$ & $0.15 \pm 0.02$ \\
			$^{16}$O & $2_1^+$ & 34   & $0.16 \pm 0.02  $ & $0.14 \pm 0.03$  \\
			\hline
			\hline
		\end{tabular}
	\end{center}
\end{table}

The differences between both approaches indicate a significant contribution from the nuclear interior. To check this point further, and following the technique of Ref.\ \cite{SP19}, we define a modified kernel $K^{J \pi}_{\alpha \beta}(r^{min}, R, R')$ in Eq. (\ref{a2}) using a cutoff radius $r^{min}$ over the internal coordinates of the projectile ($r_A$) or of the residual nucleus ($r_B$). Consequently, we have $K^{J \pi}_{\alpha \beta}(r^{min}, R, R')$ = $K^{J \pi}_{\alpha \beta}(R, R')$ for $r^{min}\leq r_A$ or $r_B$ and it is 0 for $r^{min} > r_A$ or $r_B$. This in turn results into a modified scattering matrix $\tilde{U}(r^{min})$, such that $\tilde{U}(0) = U$ and $\tilde{U}(\infty) = 0$. 
This technique permits to evaluate the role of short distances in the transfer cross section.

\begin{figure}[ht]
		\includegraphics[clip,width=0.4\textwidth]{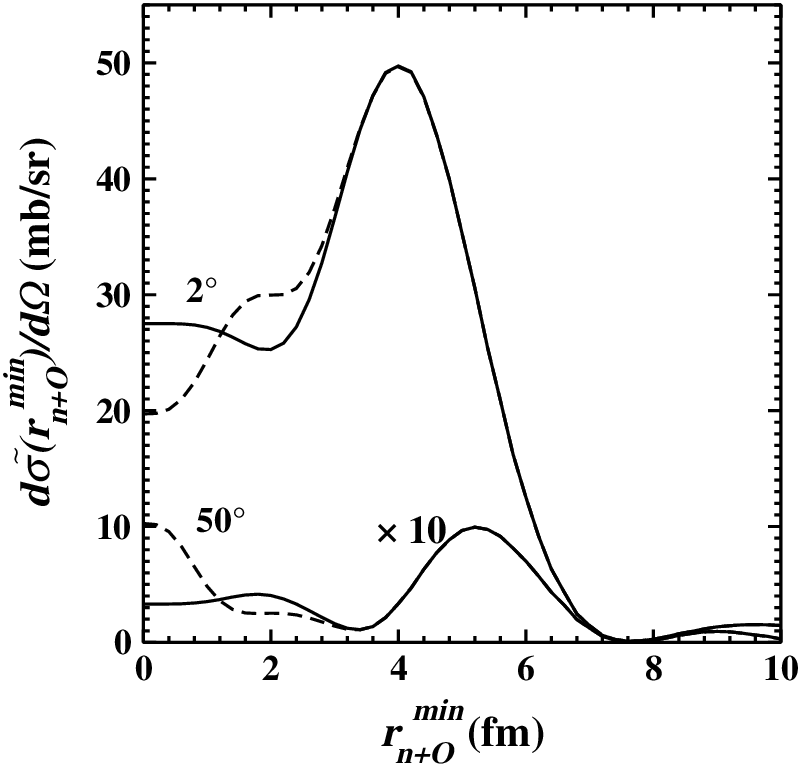}
		\caption{Modified cross sections $d\tilde{\sigma}(r^{min})/d\Omega)$ for the reaction $^{16}$O($d, p$)$^{17}$O (1/2$^+$) at $E_d = 25.4$ MeV, plotted at two different angles as a function of the cutoff distance $r^{min}$ over the $n + ^{16}$O coordinate. Dashed and solid lines have the same meaning as in Fig.\ \ref{dp_css}. } 
		\label{dp_rmin}
	\end{figure}

In Fig.\ \ref{dp_rmin}, we plot the modified cross sections $d\tilde{\sigma}(r^{min})/d\Omega$ calculated using $\tilde{U}(r^{min})$, as a function of cutoff distance $r^{min}$ over the $n + ^{16}$O coordinate. We plot these cross sections at two different scattering angles $2^\circ$ and $50^\circ$ for the deuteron energy 25.4 MeV. Dashed and solid lines are the same as in Fig. \ref{dp_css}. One can see that there is a significant difference in both these calculations for $0 \leq r^{min} \leq 3.6$ fm and this is the region where the wave functions from both these potentials [see Fig.\ \ref{wfns}(a)] used for $^{17}$O differ from each other. This clearly explains the differences between both calculations in Fig.\ \ref{dp_css}. It is also clear from Fig.\ \ref{dp_rmin} that this difference increases at large angles. Similar conclusions are drawn at $E_d = 36$ MeV. We also repeated our calculations below the Coulomb barrier ($V_C = 2.54$ MeV) using global deuteron and proton optical potentials from Refs.\ \cite{HYS06,CH89} and found that for deuteron energies well below the barrier SUSY transformations has negligible effect. This suggests that the process is completely peripheral for energies below the barrier. However, for energies far above the barrier, differences start appearing between these two calculations especially at larger angles, which further increase with the energies.


%
%

Next, we discuss the $^{12}$C($^7$Li, $t$)$^{16}$O reaction, and consider $\alpha$ transfer leading to the $0_2^+$ ($E_x = 6.05$ MeV) and $2_1^+$ ($E_x = 6.92$ MeV) states of $^{16}$O. In the literature, the $^{12}$C($^7$Li, $t$)$^{16}$O reaction has been used in indirect measurements of $^{12}$C($\alpha, \gamma$)$^{16}$O \cite{OHR12,BFH78}, which is very important 
in nuclear astrophysics. At typical stellar energies, around 300 keV, the cross section estimates rely on extrapolation procedures, due to the difficulties in the direct measurements \cite{DGW17}. In this procedure, the phenomenological R-matrix method is often adopted for fits to the various available data, and this also requires spectroscopic information about the various states of $^{16}$O. It is well known that the reduced $\alpha$ width of a bound state mainly depends upon its spectroscopic factor, and that $\alpha$ transfer reactions such as ($^7$Li, $t$) or ($^6$Li, $d$) provide an efficient way to determine them. 

\begin{figure}[ht]
	\includegraphics[clip,width=0.45\textwidth]{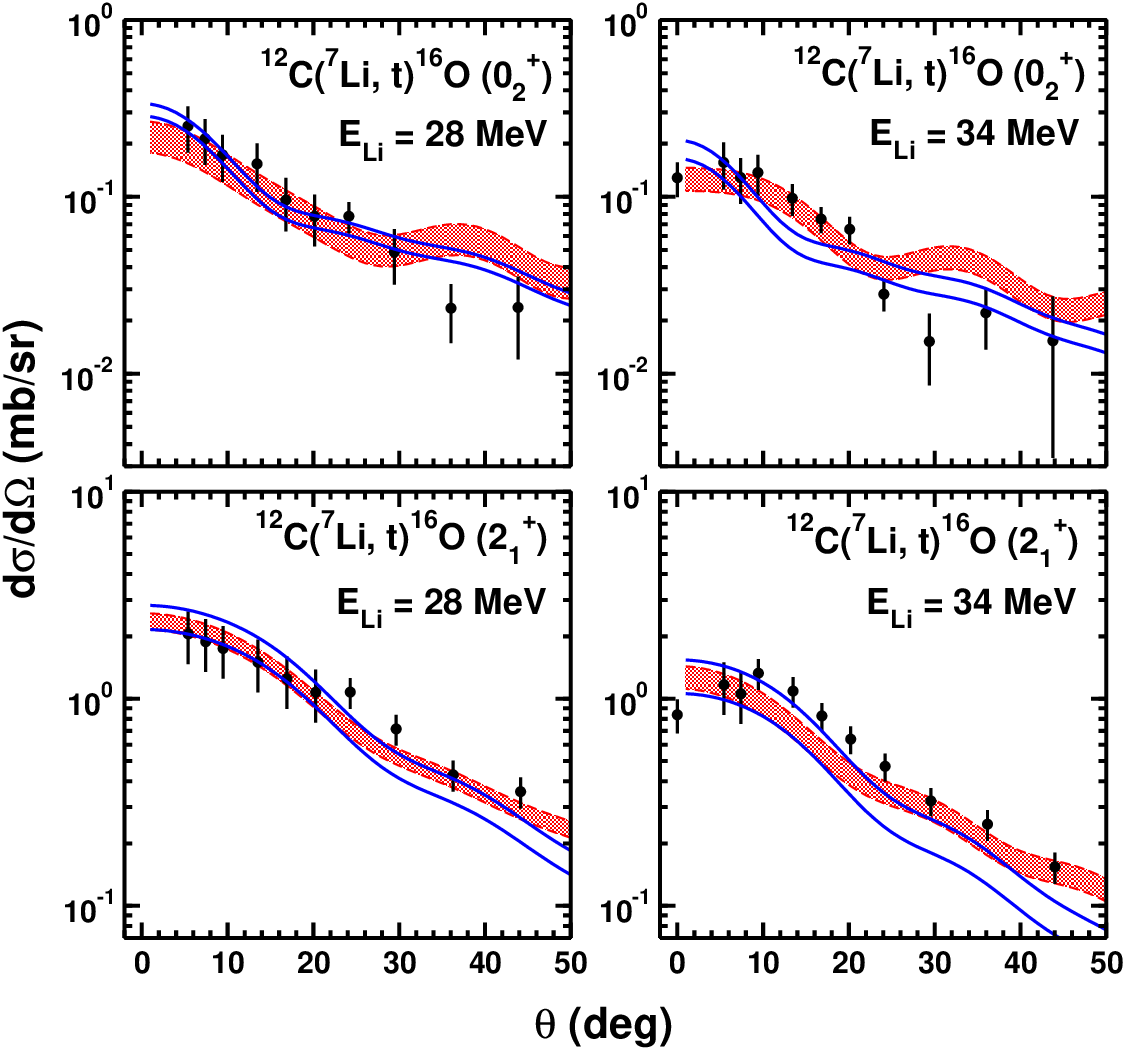}
	\caption{Angular distributions of $^{12}$C($^7$Li, $t$)$^{16}$O at two $^7$Li energies. Dashed and solid lines represent calculations using WS and SUSY bound-state potentials in both channels, respectively. Experimental data are taken from Ref.\ \cite{OHR12}. Spectroscopic factors are given in Table \ref{tab2}. The upper
	and lower limits, and the hatched regions, represent the uncertainties associated with the angular range (see text). } 
	\label{at_css}
\end{figure}

We study the $^{12}$C($^7$Li, $t$)$^{16}$O reaction at two  $^7$Li energies (28 and 34 MeV). As given in Table \ref{tab1}, there are forbidden states both in the $\alpha + t$ and in the $\alpha + ^{12}$C potentials. 
In Fig.\ \ref{at_css}, we plot the $^{12}$C($^7$Li, $t$)$^{16}$O angular distribution at 28 and 34 MeV for both $^{16}$O states. Dashed and solid lines are calculations with the WS and SUSY potentials in both channels, respectively and they are fitted to the experimental data of Ref.\ \cite{OHR12}. As there is no clear minima in the data unlike the previous case, to extract the SFs we fit all the available data points. We fit the data in steps with an increment of $10^\circ$ and extract respective SFs in each interval. The final SF for a given state is then obtained by taking the average of the SFs obtained from different intervals. In Fig.\ \ref{at_css} the upper
	and lower limits for the solid lines, and the hatched regions for the dashed lines, represent the uncertainties associated with the angular range. These $\alpha-$SFs are presented in Table \ref{tab2}. Additionally, around 46\% and 33\% uncertainties are also expected in the extracted SFs of $0_2^+$ and $2_1^+$ states, respectively, due to the ambiguities in the optical potentials and due to variation of the bound state potential parameters as mentioned in Ref. \cite{OHR12}.

The SFs with deep potentials are somewhat larger than those obtained in Ref.\ \cite{OHR12}, where they were reported as $0.13^{+0.07}_{-0.06}$ and $0.15 \pm 0.05$, respectively. This is due to the inclusion of remnant terms in our case, as it was also pointed out in Ref.\ \cite{SP19}. 
These SFs remain almost unaltered when we replace the WS potentials by their SUSY partners in both channels. However, the shapes of the angular distributions (see Fig. \ref{at_css}) are changed significantly, especially for the $0_2^+$ state.
To trace its origin, we analyze the cross sections as a function of the cutoff distance $r^{min}$ over the $\alpha-t$ and $\alpha-{\rm C}$ distances.

\begin{figure}[ht]
	\includegraphics[clip,width=0.4\textwidth]{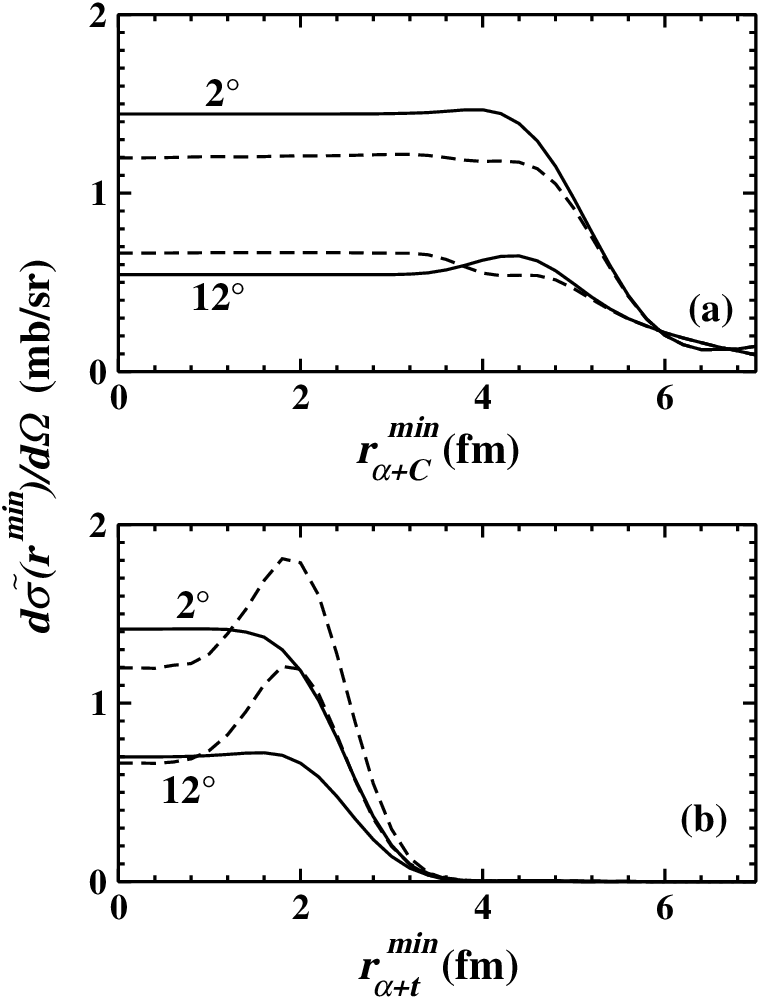}
	\caption{Modified cross sections $d\tilde{\sigma}(r^{min})/d\Omega)$ for the reaction $^{12}$C($^7$Li, $t$)$^{16}$O ($0_2^+$) at $E_{^7Li} = 28$ MeV, plotted at two different angles, as a function of cutoff distance $r^{min}$ over (a) $\alpha + ^{12}$C  and (b) $\alpha + t$ coordinates. The dashed lines have the same meaning as in Fig.\ \ref{at_css} but with SFs = 1. The solid lines represent the calculations using a SUSY potential only for $\alpha+^{12}$C (a) or for $\alpha+t$ (b). } 
	\label{at_rmin}
\end{figure}

In Fig.\ \ref{at_rmin}, we plot the modified cross section $d\tilde{\sigma}(r^{min})/d\Omega)$ at $\theta = 2^\circ$ and $ 12^\circ$ for the reaction $^{12}$C($^7$Li, $t$)$^{16}$O ($0_2^+$) at $E_{^7Li} = 28$ MeV, as a function of cutoff distance $r^{min}$ over the $\alpha + ^{12}$C (upper panel) and $\alpha + t$ (lower panel) coordinates.
Dashed lines in the figure correspond to the calculations when using WS bound state potentials in both the channels, whereas solid lines are the calculations when using SUSY potential only for the (a) $^{16}$O or for the (b) $^7$Li nucleus.  It is clear from the figure that cross sections are sensitive to the larger $\alpha + ^{12}$C distances as compared to the $\alpha + t$ distances. Furthermore, at very forward angles cross sections are equally sensitive to both these coordinates but at larger angles where shapes of angular distribution are mostly effected, cross sections are mainly governed by the $\alpha + ^{12}$C distances [see around ($r^{min}\approx 0$)].
A similar behavior was found at 34 MeV, and for the $2^+_1$ state of $^{16}$O. As we did in the previous case, here also we repeated our calculations at lower projectile energies and found that below the Coulomb barrier ($V_C = 5.1$ MeV) the process is completely peripheral in this case too with respect to both the $\alpha + t$ and $\alpha + ^{12}$C distances, whereas as one goes somewhat above the barrier nuclear interior start contributing. 

In conclusion, we have addressed the sensitivity of the DWBA cross sections against two models of bound-state wave functions.
Starting from deep WS potentials, we have built equivalent supersymmetric partners. The asymptotic part of the wave function is identical, but there are differences in the internal part.
The comparison of these calculations allow us to test the sensitivity of transfer cross sections to the nuclear interior. We have applied the model to $^{16}$O($d, p$)$^{17}$O and $^{12}$C($^7$Li, $t$)$^{16}$O, typical cases of nucleon and $\alpha$ transfer, respectively. We found a significant contribution from the nuclear interior for energies $> 15-20$ MeV. Apart from the magnitude, the shape of the angular distributions is also effected. 
For the $^{16}$O($d, p$)$^{17}$O reaction, the SFs are reduced by about 30\% when using SUSY potentials. For the
$^{12}$C($^7$Li, $t$)$^{16}$O reaction, however, the main difference between both potentials is in the shape of the
angular distribution. We have considered various limits for the angular ranges where the theory is compared to experiment. 
This provides additional uncertainties of $10-20\%$ in the spectroscopic factors apart from those due the ambiguities in the optical potentials.

\section*{Acknowledgments}
This work has received funding from the European Union's Horizon 2020 research and innovation program under the Marie Sk{\l}odowska-Curie grant agreement No 801505. It was also supported by the Fonds de la Recherche Scientifique - FNRS under Grant Numbers 4.45.10.08 and J.0049.19.







\begin{thebibliography}{45}
\expandafter\ifx\csname natexlab\endcsname\relax\def\natexlab#1{#1}\fi
\providecommand{\bibinfo}[2]{#2}
\ifx\xfnm\relax \def\xfnm[#1]{\unskip,\space#1}\fi
\bibitem[{Satchler(1990)}]{Sa90}
\bibinfo{author}{G.~Satchler}, \bibinfo{title}{Introduction to Nuclear
  Reactions}, \bibinfo{publisher}{Oxford University Press, USA},
  \bibinfo{year}{1990}.
\bibitem[{Glendenning(2004)}]{Gl04}
\bibinfo{author}{N.~Glendenning}, \bibinfo{title}{Direct Nuclear Reactions},
  \bibinfo{publisher}{World Scientific, Singapore}, \bibinfo{year}{2004}.
\bibitem[{Ohmura et~al.(1969)Ohmura, Imanishi, Ichimura, and Kawai}]{OII69}
\bibinfo{author}{T.~Ohmura}, \bibinfo{author}{B.~Imanishi},
  \bibinfo{author}{M.~Ichimura}, \bibinfo{author}{M.~Kawai},
  \bibinfo{journal}{Progress of Theoretical Physics} \bibinfo{volume}{41}
  (\bibinfo{year}{1969}) \bibinfo{pages}{391--418}.
\bibitem[{Thompson(1988)}]{Th88}
\bibinfo{author}{I.~J. Thompson}, \bibinfo{journal}{Comput. Phys. Rep.}
  \bibinfo{volume}{7} (\bibinfo{year}{1988}) \bibinfo{pages}{167}.
\bibitem[{Thomas et~al.(2005)Thomas, Bardayan, Blackmon, Cizewski, Greife,
  Gross, Johnson, Jones, Kozub, Liang, Livesay, Ma, Moazen, Nesaraja, Shapira,
  and Smith}]{THO05}
\bibinfo{author}{J.~S. Thomas}, \bibinfo{author}{D.~W. Bardayan},
  \bibinfo{author}{J.~C. Blackmon}, \bibinfo{author}{J.~A. Cizewski},
  \bibinfo{author}{U.~Greife}, \bibinfo{author}{C.~J. Gross},
  \bibinfo{author}{M.~S. Johnson}, \bibinfo{author}{K.~L. Jones},
  \bibinfo{author}{R.~L. Kozub}, \bibinfo{author}{J.~F. Liang},
  \bibinfo{author}{R.~J. Livesay}, \bibinfo{author}{Z.~Ma},
  \bibinfo{author}{B.~H. Moazen}, \bibinfo{author}{C.~D. Nesaraja},
  \bibinfo{author}{D.~Shapira}, \bibinfo{author}{M.~S. Smith},
  \bibinfo{journal}{Phys. Rev. C} \bibinfo{volume}{71} (\bibinfo{year}{2005})
  \bibinfo{pages}{021302}.
\bibitem[{Timofeyuk and Johnson(2020)}]{TJ20}
\bibinfo{author}{N.~Timofeyuk}, \bibinfo{author}{R.~Johnson},
  \bibinfo{journal}{Prog. Part. Nucl. Phys.} \bibinfo{volume}{111}
  (\bibinfo{year}{2020}) \bibinfo{pages}{103738}.
\bibitem[{Bardayan(2016)}]{BAR16}
\bibinfo{author}{D.~W. Bardayan}, \bibinfo{journal}{J. Phys. G: Nucl. Part.
  Phys.} \bibinfo{volume}{43} (\bibinfo{year}{2016}) \bibinfo{pages}{043001}.
\bibitem[{Bertulani et~al.(2016)Bertulani, Shubhchintak, Mukhamedzhanov,
  Kadyrov, Kruppa, and Pang}]{BER16}
\bibinfo{author}{C.~Bertulani}, \bibinfo{author}{Shubhchintak},
  \bibinfo{author}{A.~Mukhamedzhanov}, \bibinfo{author}{A.~S. Kadyrov},
  \bibinfo{author}{A.~Kruppa}, \bibinfo{author}{D.~Y. Pang},
  \bibinfo{journal}{J. Phys.: Conf. Ser.} \bibinfo{volume}{703}
  (\bibinfo{year}{2016}) \bibinfo{pages}{012007}.
\bibitem[{Tribble et~al.(2014)Tribble, Bertulani, Cognata, Mukhamedzhanov, and
  Spitaleri}]{TRI14}
\bibinfo{author}{R.~E. Tribble}, \bibinfo{author}{C.~A. Bertulani},
  \bibinfo{author}{M.~L. Cognata}, \bibinfo{author}{A.~M. Mukhamedzhanov},
  \bibinfo{author}{C.~Spitaleri}, \bibinfo{journal}{Rep. Prog. Phys.}
  \bibinfo{volume}{77} (\bibinfo{year}{2014}) \bibinfo{pages}{106901}.
\bibitem[{Mukhamedzhanov et~al.(2008)Mukhamedzhanov, Nunes, and Mohr}]{AKR08}
\bibinfo{author}{A.~M. Mukhamedzhanov}, \bibinfo{author}{F.~M. Nunes},
  \bibinfo{author}{P.~Mohr}, \bibinfo{journal}{Phys. Rev. C}
  \bibinfo{volume}{77} (\bibinfo{year}{2008}) \bibinfo{pages}{051601}.
\bibitem[{Bertulani et~al.(2019)Bertulani, Canto, Hussein, Shubhchintak, and
  Nhan~Hao}]{BER19}
\bibinfo{author}{C.~A. Bertulani}, \bibinfo{author}{L.~F. Canto},
  \bibinfo{author}{M.~S. Hussein}, \bibinfo{author}{Shubhchintak},
  \bibinfo{author}{T.~V. Nhan~Hao}, \bibinfo{journal}{Int. J. Mod. Phys. E}
  \bibinfo{volume}{28} (\bibinfo{year}{2019}) \bibinfo{pages}{1950109}.
\bibitem[{Satchler(1983)}]{Sa83}
\bibinfo{author}{G.~R. Satchler}, \bibinfo{title}{Direct Nuclear Reactions},
  \bibinfo{publisher}{Oxford University Press}, \bibinfo{year}{1983}.
\bibitem[{Johnson and Soper(1970)}]{JS70}
\bibinfo{author}{R.~C. Johnson}, \bibinfo{author}{P.~J.~R. Soper},
  \bibinfo{journal}{Phys. Rev. C} \bibinfo{volume}{1} (\bibinfo{year}{1970})
  \bibinfo{pages}{976}.
\bibitem[{Austern et~al.(1987)Austern, Iseri, Kamimura, Kawai, Rawitscher, and
  Yahiro}]{AIK87}
\bibinfo{author}{N.~Austern}, \bibinfo{author}{Y.~Iseri},
  \bibinfo{author}{M.~Kamimura}, \bibinfo{author}{M.~Kawai},
  \bibinfo{author}{G.~Rawitscher}, \bibinfo{author}{M.~Yahiro},
  \bibinfo{journal}{Phys. Rep.} \bibinfo{volume}{154} (\bibinfo{year}{1987})
  \bibinfo{pages}{125}.
\bibitem[{Deltuva(2013)}]{De13}
\bibinfo{author}{A.~Deltuva}, \bibinfo{journal}{Phys. Rev. C}
  \bibinfo{volume}{88} (\bibinfo{year}{2013}) \bibinfo{pages}{011601}.
\bibitem[{Moro et~al.(2009)Moro, Nunes, and Johnson}]{MNJ09}
\bibinfo{author}{A.~M. Moro}, \bibinfo{author}{F.~M. Nunes},
  \bibinfo{author}{R.~C. Johnson}, \bibinfo{journal}{Phys. Rev. C}
  \bibinfo{volume}{80} (\bibinfo{year}{2009}) \bibinfo{pages}{064606}.
\bibitem[{Nguyen et~al.(2010)Nguyen, Nunes, and Johnson}]{NNJ10}
\bibinfo{author}{N.~B. Nguyen}, \bibinfo{author}{F.~M. Nunes},
  \bibinfo{author}{R.~C. Johnson}, \bibinfo{journal}{Phys. Rev. C}
  \bibinfo{volume}{82} (\bibinfo{year}{2010}) \bibinfo{pages}{014611}.
\bibitem[{Schmitt et~al.(2013)Schmitt, Jones, Ahn, Bardayan, Bey, Blackmon,
  Brown, Chae, Chipps, Cizewski, Hahn, Kolata, Kozub, Liang, Matei, Matos,
  Matyas, Moazen, Nesaraja, Nunes, O'Malley, Pain, Peters, Pittman, Roberts,
  Shapira, Shriner, Smith, Spassova, Stracener, Upadhyay, Villano, and
  Wilson}]{SJA13}
\bibinfo{author}{K.~T. Schmitt}, \bibinfo{author}{K.~L. Jones},
  \bibinfo{author}{S.~Ahn}, \bibinfo{author}{D.~W. Bardayan},
  \bibinfo{author}{A.~Bey}, \bibinfo{author}{J.~C. Blackmon},
  \bibinfo{author}{S.~M. Brown}, \bibinfo{author}{K.~Y. Chae},
  \bibinfo{author}{K.~A. Chipps}, \bibinfo{author}{J.~A. Cizewski},
  \bibinfo{author}{K.~I. Hahn}, \bibinfo{author}{J.~J. Kolata},
  \bibinfo{author}{R.~L. Kozub}, \bibinfo{author}{J.~F. Liang},
  \bibinfo{author}{C.~Matei}, \bibinfo{author}{M.~Matos},
  \bibinfo{author}{D.~Matyas}, \bibinfo{author}{B.~Moazen},
  \bibinfo{author}{C.~D. Nesaraja}, \bibinfo{author}{F.~M. Nunes},
  \bibinfo{author}{P.~D. O'Malley}, \bibinfo{author}{S.~D. Pain},
  \bibinfo{author}{W.~A. Peters}, \bibinfo{author}{S.~T. Pittman},
  \bibinfo{author}{A.~Roberts}, \bibinfo{author}{D.~Shapira},
  \bibinfo{author}{J.~F. Shriner}, \bibinfo{author}{M.~S. Smith},
  \bibinfo{author}{I.~Spassova}, \bibinfo{author}{D.~W. Stracener},
  \bibinfo{author}{N.~J. Upadhyay}, \bibinfo{author}{A.~N. Villano},
  \bibinfo{author}{G.~L. Wilson}, \bibinfo{journal}{Phys. Rev. C}
  \bibinfo{volume}{88} (\bibinfo{year}{2013}) \bibinfo{pages}{064612}.
\bibitem[{Walter et~al.(2019)Walter, Pain, Cizewski, Nunes, Ahn, Baugher,
  Bardayan, Baumann, Bazin, Burcher, Chipps, Cerizza, Jones, Kozub, Lonsdale,
  Manning, Montes, O'Malley, Ota, Pereira, Ratkiewicz, Thompson, Thornsberry,
  and Williams}]{WPC19}
\bibinfo{author}{D.~Walter}, \bibinfo{author}{S.~D. Pain},
  \bibinfo{author}{J.~A. Cizewski}, \bibinfo{author}{F.~M. Nunes},
  \bibinfo{author}{S.~Ahn}, \bibinfo{author}{T.~Baugher},
  \bibinfo{author}{D.~W. Bardayan}, \bibinfo{author}{T.~Baumann},
  \bibinfo{author}{D.~Bazin}, \bibinfo{author}{S.~Burcher},
  \bibinfo{author}{K.~A. Chipps}, \bibinfo{author}{G.~Cerizza},
  \bibinfo{author}{K.~L. Jones}, \bibinfo{author}{R.~L. Kozub},
  \bibinfo{author}{S.~J. Lonsdale}, \bibinfo{author}{B.~Manning},
  \bibinfo{author}{F.~Montes}, \bibinfo{author}{P.~D. O'Malley},
  \bibinfo{author}{S.~Ota}, \bibinfo{author}{J.~Pereira},
  \bibinfo{author}{A.~Ratkiewicz}, \bibinfo{author}{P.~Thompson},
  \bibinfo{author}{C.~Thornsberry}, \bibinfo{author}{S.~Williams},
  \bibinfo{journal}{Phys. Rev. C} \bibinfo{volume}{99} (\bibinfo{year}{2019})
  \bibinfo{pages}{054625}.
\bibitem[{Jayatissa et~al.(2020)Jayatissa, Rogachev, Goldberg, Koshchiy,
  Christian, Hooker, Ota, Roeder, Saastamoinen, Trippella, Upadhyayula, and
  Uberseder}]{JRG20}
\bibinfo{author}{H.~Jayatissa}, \bibinfo{author}{G.~Rogachev},
  \bibinfo{author}{V.~Goldberg}, \bibinfo{author}{E.~Koshchiy},
  \bibinfo{author}{G.~Christian}, \bibinfo{author}{J.~Hooker},
  \bibinfo{author}{S.~Ota}, \bibinfo{author}{B.~Roeder},
  \bibinfo{author}{A.~Saastamoinen}, \bibinfo{author}{O.~Trippella},
  \bibinfo{author}{S.~Upadhyayula}, \bibinfo{author}{E.~Uberseder},
  \bibinfo{journal}{Phys. Lett. B} \bibinfo{volume}{802} (\bibinfo{year}{2020})
  \bibinfo{pages}{135267}.
\bibitem[{Hamill et~al.(2020)Hamill, Woods, Kahl, Longland, Greene, Marshall,
  Portillo, and Setoodehnia}]{HWK20}
\bibinfo{author}{C.~B. Hamill}, \bibinfo{author}{P.~J. Woods},
  \bibinfo{author}{D.~Kahl}, \bibinfo{author}{R.~Longland},
  \bibinfo{author}{J.~P. Greene}, \bibinfo{author}{C.~Marshall},
  \bibinfo{author}{F.~Portillo}, \bibinfo{author}{K.~Setoodehnia},
  \bibinfo{journal}{Eur. Phys. J. A} \bibinfo{volume}{56}
  (\bibinfo{year}{2020}) \bibinfo{pages}{36}.
\bibitem[{King et~al.(2018)King, Lovell, and Nunes}]{KLN18}
\bibinfo{author}{G.~B. King}, \bibinfo{author}{A.~E. Lovell},
  \bibinfo{author}{F.~M. Nunes}, \bibinfo{journal}{Phys. Rev. C}
  \bibinfo{volume}{98} (\bibinfo{year}{2018}) \bibinfo{pages}{044623}.
\bibitem[{Str\"omich et~al.(1977)Str\"omich, Steinmetz, Bangert, Gonsior, Roth,
  and von Brentano}]{SSB77}
\bibinfo{author}{A.~Str\"omich}, \bibinfo{author}{B.~Steinmetz},
  \bibinfo{author}{R.~Bangert}, \bibinfo{author}{B.~Gonsior},
  \bibinfo{author}{M.~Roth}, \bibinfo{author}{P.~von Brentano},
  \bibinfo{journal}{Phys. Rev. C} \bibinfo{volume}{16} (\bibinfo{year}{1977})
  \bibinfo{pages}{2193}.
\bibitem[{Schmitt et~al.(2012)Schmitt, Jones, Bey, Ahn, Bardayan, Blackmon,
  Brown, Chae, Chipps, Cizewski, Hahn, Kolata, Kozub, Liang, Matei,
  Mato\ifmmode~\check{s}\else \v{s}\fi{}, Matyas, Moazen, Nesaraja, Nunes,
  O'Malley, Pain, Peters, Pittman, Roberts, Shapira, Shriner, Smith, Spassova,
  Stracener, Villano, and Wilson}]{SJB12}
\bibinfo{author}{K.~T. Schmitt}, \bibinfo{author}{K.~L. Jones},
  \bibinfo{author}{A.~Bey}, \bibinfo{author}{S.~H. Ahn}, \bibinfo{author}{D.~W.
  Bardayan}, \bibinfo{author}{J.~C. Blackmon}, \bibinfo{author}{S.~M. Brown},
  \bibinfo{author}{K.~Y. Chae}, \bibinfo{author}{K.~A. Chipps},
  \bibinfo{author}{J.~A. Cizewski}, \bibinfo{author}{K.~I. Hahn},
  \bibinfo{author}{J.~J. Kolata}, \bibinfo{author}{R.~L. Kozub},
  \bibinfo{author}{J.~F. Liang}, \bibinfo{author}{C.~Matei},
  \bibinfo{author}{M.~Mato\ifmmode~\check{s}\else \v{s}\fi{}},
  \bibinfo{author}{D.~Matyas}, \bibinfo{author}{B.~Moazen},
  \bibinfo{author}{C.~Nesaraja}, \bibinfo{author}{F.~M. Nunes},
  \bibinfo{author}{P.~D. O'Malley}, \bibinfo{author}{S.~D. Pain},
  \bibinfo{author}{W.~A. Peters}, \bibinfo{author}{S.~T. Pittman},
  \bibinfo{author}{A.~Roberts}, \bibinfo{author}{D.~Shapira},
  \bibinfo{author}{J.~F. Shriner}, \bibinfo{author}{M.~S. Smith},
  \bibinfo{author}{I.~Spassova}, \bibinfo{author}{D.~W. Stracener},
  \bibinfo{author}{A.~N. Villano}, \bibinfo{author}{G.~L. Wilson},
  \bibinfo{journal}{Phys. Rev. Lett.} \bibinfo{volume}{108}
  (\bibinfo{year}{2012}) \bibinfo{pages}{192701}.
\bibitem[{Pang et~al.(2007)Pang, Nunes, and Mukhamedzhanov}]{PNM07}
\bibinfo{author}{D.~Y. Pang}, \bibinfo{author}{F.~M. Nunes},
  \bibinfo{author}{A.~M. Mukhamedzhanov}, \bibinfo{journal}{Phys. Rev. C}
  \bibinfo{volume}{75} (\bibinfo{year}{2007}) \bibinfo{pages}{024601}.
\bibitem[{Yang and Capel(2018)}]{YC18}
\bibinfo{author}{J.~Yang}, \bibinfo{author}{P.~Capel}, \bibinfo{journal}{Phys.
  Rev. C} \bibinfo{volume}{98} (\bibinfo{year}{2018}) \bibinfo{pages}{054602}.
\bibitem[{Sukumar(1985)}]{SUK85}
\bibinfo{author}{C.~V. Sukumar}, \bibinfo{journal}{J. Phys. A: Math. Gen.}
  \bibinfo{volume}{18} (\bibinfo{year}{1985}) \bibinfo{pages}{2917}.
\bibitem[{Baye(1987)}]{BAY87}
\bibinfo{author}{D.~Baye}, \bibinfo{journal}{Phys. Rev. Lett.}
  \bibinfo{volume}{58} (\bibinfo{year}{1987}) \bibinfo{pages}{2738}.
\bibitem[{Ridikas et~al.(1996)Ridikas, Vaagen, and Bang}]{RVB96}
\bibinfo{author}{D.~Ridikas}, \bibinfo{author}{J.~Vaagen},
  \bibinfo{author}{J.~Bang}, \bibinfo{journal}{Nucl. Phys. A}
  \bibinfo{volume}{609} (\bibinfo{year}{1996}) \bibinfo{pages}{21}.
\bibitem[{Shubhchintak and Descouvemont(2019)}]{SP19}
\bibinfo{author}{Shubhchintak}, \bibinfo{author}{P.~Descouvemont},
  \bibinfo{journal}{Phys. Rev. C} \bibinfo{volume}{100} (\bibinfo{year}{2019})
  \bibinfo{pages}{034611}.
\bibitem[{Descouvemont and Baye(2010)}]{PDB10}
\bibinfo{author}{P.~Descouvemont}, \bibinfo{author}{D.~Baye},
  \bibinfo{journal}{Rep. Prog. Phys.} \bibinfo{volume}{73}
  (\bibinfo{year}{2010}) \bibinfo{pages}{036301}.
\bibitem[{Baye(2015)}]{BAYE15}
\bibinfo{author}{D.~Baye}, \bibinfo{journal}{Phys. Rep.} \bibinfo{volume}{565}
  (\bibinfo{year}{2015}) \bibinfo{pages}{1}.
\bibitem[{Thompson and Nunes(2009)}]{TN09}
\bibinfo{author}{I.~Thompson}, \bibinfo{author}{F.~Nunes},
  \bibinfo{title}{Nuclear Reactions for Astrophysics: Principles, Calculation
  and Applications of Low-Energy Reactions}, \bibinfo{publisher}{Cambridge
  University Press}, \bibinfo{year}{2009}.
\bibitem[{Cooper et~al.(1974)Cooper, Hornyak, and Roos}]{CHR74}
\bibinfo{author}{M.~D. Cooper}, \bibinfo{author}{W.~F. Hornyak},
  \bibinfo{author}{P.~G. Roos}, \bibinfo{journal}{Nucl. Phys. A}
  \bibinfo{volume}{218} (\bibinfo{year}{1974}) \bibinfo{pages}{249}.
\bibitem[{Oulebsir et~al.(2012)Oulebsir, Hammache, Roussel, Pellegriti,
  Audouin, Beaumel, Bouda, Descouvemont, Fortier, Gaudefroy, Kiener,
  Lefebvre-Schuhl, and Tatischeff}]{OHR12}
\bibinfo{author}{N.~Oulebsir}, \bibinfo{author}{F.~Hammache},
  \bibinfo{author}{P.~Roussel}, \bibinfo{author}{M.~G. Pellegriti},
  \bibinfo{author}{L.~Audouin}, \bibinfo{author}{D.~Beaumel},
  \bibinfo{author}{A.~Bouda}, \bibinfo{author}{P.~Descouvemont},
  \bibinfo{author}{S.~Fortier}, \bibinfo{author}{L.~Gaudefroy},
  \bibinfo{author}{J.~Kiener}, \bibinfo{author}{A.~Lefebvre-Schuhl},
  \bibinfo{author}{V.~Tatischeff}, \bibinfo{journal}{Phys. Rev. C}
  \bibinfo{volume}{85} (\bibinfo{year}{2012}) \bibinfo{pages}{035804}.
\bibitem[{Pang et~al.(2015)Pang, Dean, and Mukhamedzhanov}]{PDM15}
\bibinfo{author}{D.~Y. Pang}, \bibinfo{author}{W.~M. Dean},
  \bibinfo{author}{A.~M. Mukhamedzhanov}, \bibinfo{journal}{Phys. Rev. C}
  \bibinfo{volume}{91} (\bibinfo{year}{2015}) \bibinfo{pages}{024611}.
\bibitem[{Koning and Delaroche(2003)}]{KD03}
\bibinfo{author}{A.~J. Koning}, \bibinfo{author}{J.~P. Delaroche},
  \bibinfo{journal}{Nucl. Phys. A} \bibinfo{volume}{713} (\bibinfo{year}{2003})
  \bibinfo{pages}{231}.
\bibitem[{Schumacher et~al.(1973)Schumacher, Ueta, Duhm, Kubo, and
  Klages}]{SUD73}
\bibinfo{author}{P.~Schumacher}, \bibinfo{author}{N.~Ueta},
  \bibinfo{author}{H.~Duhm}, \bibinfo{author}{K.-I. Kubo},
  \bibinfo{author}{W.~Klages}, \bibinfo{journal}{Nucl. Phys. A}
  \bibinfo{volume}{212} (\bibinfo{year}{1973}) \bibinfo{pages}{573 -- 599}.
\bibitem[{Garrett and Hansen(1973)}]{GH73}
\bibinfo{author}{J.~Garrett}, \bibinfo{author}{O.~Hansen},
  \bibinfo{journal}{Nucl. Phys. A} \bibinfo{volume}{212} (\bibinfo{year}{1973})
  \bibinfo{pages}{600}.
\bibitem[{Liu et~al.(2004)Liu, Famiano, Lynch, Tsang, and Tostevin}]{LFL04}
\bibinfo{author}{X.~D. Liu}, \bibinfo{author}{M.~A. Famiano},
  \bibinfo{author}{W.~G. Lynch}, \bibinfo{author}{M.~B. Tsang},
  \bibinfo{author}{J.~A. Tostevin}, \bibinfo{journal}{Phys. Rev. C}
  \bibinfo{volume}{69} (\bibinfo{year}{2004}) \bibinfo{pages}{064313}.
\bibitem[{Baye and Timofeyuk(1992)}]{BT92}
\bibinfo{author}{D.~Baye}, \bibinfo{author}{N.~K. Timofeyuk},
  \bibinfo{journal}{Phys. Lett. B} \bibinfo{volume}{293} (\bibinfo{year}{1992})
  \bibinfo{pages}{13}.
\bibitem[{Han et~al.(2006)Han, Shi, and Shen}]{HYS06}
\bibinfo{author}{Y.~Han}, \bibinfo{author}{Y.~Shi}, \bibinfo{author}{Q.~Shen},
  \bibinfo{journal}{Phys. Rev. C} \bibinfo{volume}{74} (\bibinfo{year}{2006})
  \bibinfo{pages}{044615}.
\bibitem[{Varner et~al.(1991)Varner, Thompson, McAbee, Ludwig, and
  Clegg}]{CH89}
\bibinfo{author}{R.~Varner}, \bibinfo{author}{W.~Thompson},
  \bibinfo{author}{T.~McAbee}, \bibinfo{author}{E.~Ludwig},
  \bibinfo{author}{T.~Clegg}, \bibinfo{journal}{Phys. Rep.}
  \bibinfo{volume}{201} (\bibinfo{year}{1991}) \bibinfo{pages}{57}.
\bibitem[{Becchetti et~al.(1978)Becchetti, Flynn, Hanson, and Sunier}]{BFH78}
\bibinfo{author}{F.~D. Becchetti}, \bibinfo{author}{E.~R. Flynn},
  \bibinfo{author}{D.~L. Hanson}, \bibinfo{author}{J.~W. Sunier},
  \bibinfo{journal}{Nucl. Phys. A} \bibinfo{volume}{305} (\bibinfo{year}{1978})
  \bibinfo{pages}{293}.
\bibitem[{deBoer et~al.(2017)deBoer, G\"orres, Wiescher, Azuma, Best, Brune,
  Fields, Jones, Pignatari, Sayre, Smith, Timmes, and Uberseder}]{DGW17}
\bibinfo{author}{R.~J. deBoer}, \bibinfo{author}{J.~G\"orres},
  \bibinfo{author}{M.~Wiescher}, \bibinfo{author}{R.~E. Azuma},
  \bibinfo{author}{A.~Best}, \bibinfo{author}{C.~R. Brune},
  \bibinfo{author}{C.~E. Fields}, \bibinfo{author}{S.~Jones},
  \bibinfo{author}{M.~Pignatari}, \bibinfo{author}{D.~Sayre},
  \bibinfo{author}{K.~Smith}, \bibinfo{author}{F.~X. Timmes},
  \bibinfo{author}{E.~Uberseder}, \bibinfo{journal}{Rev. Mod. Phys.}
  \bibinfo{volume}{89} (\bibinfo{year}{2017}) \bibinfo{pages}{035007}.

\end{thebibliography}



\end{document}